\documentclass[jkps,preprint,fleqn]{revtex4} 
\usepackage{graphicx}
\usepackage{amssymb}
\usepackage{amsmath}
\usepackage{bm}
\usepackage{changepage} 

\pdfoutput=1

\usepackage{float} \usepackage{graphicx}
\usepackage{epstopdf}
\usepackage{graphics}
\usepackage{caption}
\usepackage{subfigure}
\usepackage{epsfig}
\usepackage{color}
\usepackage{multirow,array}

\usepackage{tensor}

\usepackage{adjustbox}  

\usepackage{tabularray}

\usepackage{hyperref}
\hypersetup{
	colorlinks=true,
	linkcolor=blue,
	filecolor=magneta,      
	urlcolor=blue,
}

\newcommand{\tm}{\tilde{m}}

\begin{document}
\setcounter{page}{0}
\title[]{Cosmography of the minimally extended Varying Speed of Light Model}
\author{Seokcheon \surname{Lee}}

\email{skylee@skku.edu}

\affiliation{Department of Physics, Institute of Basic Science, Sungkyunkwan University, Suwon 16419, Korea}

\date[]{Received }

\begin{abstract}
Cosmography, as an integral branch of cosmology, strives to characterize the Universe without relying on pre-determined cosmological models. This model-independent approach utilizes Taylor series expansions around the current epoch, providing a direct correlation with cosmological observations and the potential to constrain theoretical models. Cosmologists can describe many measurable aspects of cosmology by using various combinations of cosmographic parameters. The varying speed of light model can be naturally implemented, provided that we do not make any further assumptions from the Robertson-Walker metric for cosmological time dilation. Therefore, we apply cosmography to the so-called minimally extended varying speed of light model. In this case, other cosmographic parameters can be used to construct the Hubble parameter for both the standard model and the varying speed-of-light model. On the other hand, distinct combinations of cosmographic values for the luminosity distance indicate the two models. Hence, luminosity distance might provide a method to constrain the parameters in varying speed-of-light models.
\end{abstract}



\maketitle


\section{Introduction}

One can characterize the expansion of the Universe by the dimensionless scale factor $a(t)$. The evolution of the scale factor is determined by the Friedmann equations of general relativity, specifically for a spatially homogeneous and isotropic universe. Recent empirical research indicates an accelerated expansion of the Universe. Essentially, this implies that the second derivative of the scale factor, denoted as $\ddot  {a}(t)$, is positive, or, equivalently, that the first derivative $\dot  {a}(t)$ progressively increases over time. Moreover, theoretically, the scale factor can be Taylor expanded from the current epoch to a nearby time, and the dimensionless coefficients of these expansions are referred to as cosmographic parameters. Although the scale factor is not directly observable, it is possible to constrain cosmographic parameters based on cosmological observations. This approach, utilizing the kinematics of the scale factor, is called cosmography, allowing for the study of the cosmological evolution of the scale factor and, conversely, determining other physical quantities that govern the dynamics of the scale factor~\cite{Harrison:1976,Visser:2003vq,Visser:2004bf,Cattoen:2007id,Cattoen:2007sk,Xu:2010PLB,Luongo:2011zz,Aviles:2012ay,Bamba:2012cp,Aviles:2012ir,Busti:2015xqa,Dunsby:2015ers,Luongo:2015zgq,Zhou:2016nik,delaCruz-Dombriz:2016rxm,Bolotin:2017yom,delaCruzDombriz:2018gsj,TeppaPannia:2018ale,Bolotin:2018xtq,Yin:2019rgm,Li:2019qic,Capozziello:2019cav,Capozziello:2020ctn,Lizardo:2020wxw,Hu:2022udt,Martins:2022unf,Rocha:2022gog,Gao:2023izj,Petreca:2023nhy,Zhang:2023eup}.  

Cosmology stands as one of the most fruitful and pertinent applications of Einstein’s General Theory of Relativity (GR), offering a framework to comprehend the behavior of the Universe in its entirety. This scientific endeavor thrives by aligning its assumptions with the observable Universe. Notably, the uniformity of the temperature of the cosmic microwave background (CMB) across different sky directions provides compelling evidence for the isotropy of the Universe on its grandest scales~\cite{Hinshaw:2013,Planck:2018nkj}. Moreover, the absence of a preferred center in the Universe suggests a corresponding homogeneity, reflected in the uniform distribution of matter across scales exceeding 100 million light years~\cite{Guzzo:2018xbe,DES:2020sjz}.  These observations lead to the formulation of the Cosmological Principle (CP), which asserts that the Universe exhibits uniformity from any vantage point in space and that all directions hold equal significance. Hence, adopting the standard form of the Robertson--Walker (RW) metric in cosmology for a cosmic time t is a viable approach~\cite{Robertson:1929,Robertson:1933,Walker:1935,Walker:1937}.

The Lorentz transformation (LT) elucidates the implications of special relativity (SR) when transitioning between two Galilean frames (GFs). In the realm of the general theory of relativity (GR), an inertial frame (IF) pertains to one that is in free fall. One can establish a Lorentz-invariant spacetime interval by comparing the coordinate disparities between two events, with this interval being deemed light-like (or null) if it registers zero. One can regard such occurrences as signals traversing Minkowski spacetime at the speed of light, bridging events separated by the null interval~\cite{Morin07}. However, within the context of GR, the delineation of a global time proves elusive due to the absence of a universal inertial frame. Nonetheless, under specific conditions and with a metric adhering to the cosmological principle, a global time for the Universe can be defined. It is achieved by foliating spacetime into a series of non-intersecting space-like 3D surfaces~\cite{Weyl:1923,Islam02,Narlikar02,Hobson06,Gron07,Ryder09,CB15,Roos15,Guidry19,Ferrari21}.

In SR, the sole parameter is the speed of light in a vacuum, denoted as c. Through our research, we have demonstrated that adhering to the universal Lorentz covariance or adopting the single postulate of Minkowski spacetime, is sufficient to satisfy the principles of SR~\cite{Das93,Schutz97,Lee:2020zts}.  Consequently, this allows for the formulation of models such as the Lorentz invariant (LI) varying speed of light (VSL) model, provided that c remains locally constant while allowing for changes at cosmological scales. However, to prevent trivial alterations in units, one needs to examine the simultaneous variation in both c and Newton’s gravitational constant, G. This is essential as c and G are combined as $G/c^4$ in the Einstein action~\cite{Barrow:1998eh}.

All galaxies are presumed to occupy a hypersurface, aligning the surface of simultaneity with a local Lorentz frame (LF) of each nearby galaxy. From this arrangement, it can be deduced that a universal cosmic time exists for all galaxies situated on the hypersurface, contingent upon the homogeneity and isotropy of space, which implies a constant spatial curvature. Cosmic time denotes the measurement observed by a comoving observer who perceives the Universe's uniform expansion around them, where light traverses through the expanding space. The manifestation of a cosmological redshift in sufficiently distant light sources, corresponding to their distance from us, follows Hubble’s law, an observation induced by the CP.

In addressing problematic observational outcomes within the framework of GR, the notion of VSL has been occasionally invoked. Einstein postulated that a decrease in the wavelength $\lambda$ corresponds to a decrease in the speed of light, expressed as $c = \nu \lambda$ with a constant frequency $\nu$, and suggested that a gravitational field induces a time dilation (TD) effect represented by $\nu_1 = \nu_2 (1 + GM/rc^2)$~\cite{Einstein:1911}. Dicke introduced the concept of varying wavelength and frequency by defining a refractive index $n \equiv c/c_0 = 1 + 2GM/rc^2$~\cite{Dicke:1957}, proposing an alternative cosmological description to account for the cosmological redshift by incorporating a time-varying $c$. These seminal works hypothesized that TD arises from local gravity. Various cosmologically motivated VSL models have been developed to address issues such as the horizon problem of the Big Bang model and offer an alternative approach to cosmic inflation~\cite{Barrow:1998eh,Petit:1988,Petit:1988-2,Petit:1989,Midy:1989,Moffat:1992ud,Petit:1995ass,Albrecht:1998ir,Barrow:1998he,Clayton:1998hv,Barrow:1999jq,Clayton:1999zs,Brandenberger:1999bi,Bassett:2000wj,Gopakumar:2000kp,Magueijo:2000zt,Magueijo:2000au,Magueijo:2003gj,Magueijo:2007gf,Petit:2008eb,Roshan:2009yb,Sanejouand:2009,Nassif:2012dr,Moffat:2014poa,Ravanpak:2017kdg,Costa:2017abc,Nassif:2018pdu}. The minimal VSL (mVSL) model posits changes solely in the speed of light while keeping other physical constants constant. Petit argued that if $c$ varies over cosmic time, concurrent variations in all related physical constants should be considered, ensuring consistency with physical laws throughout the Universe's evolution. This approach seeks to establish a universal gauge relationship and temporal variation in constants while upholding the consistency of physical equations and measurements~\cite{Petit:1995ass,Petit:2008eb}:
\begin{align}
G &= G_0 a^{-1} \,, \quad m = m_{0} a  \,, \quad c = c_0 a^{\frac{1}{2}}  \,, \quad h = h_0 a^{\frac{3}{2}} \,, \quad e = e_{0} a^{\frac{1}{2}}  \,, \quad \mu = \mu_0 a \label{Petitconst} \,.
\end{align}

It 
 is important to underscore that cosmology-driven VSL models, as distinct from those rooted in local gravity effects such as~\cite{Einstein:1911,Dicke:1957}, attribute cosmic TD to both the Universe's expansion and variations in the local speed of light within the framework of the RW metric~\cite{Barrow:1998he,Lee:2023rqv}.

In the conventional RW metric, the assumption of constant light speed hinges on a specific hypothesis concerning cosmological TD rather than directly emerging from the metric's foundational principles~\cite{Lee:2023rqv,Lee:2023ucu,Lee:2023FoP,SLee:2024}. It has been shown that the possibility of the VSL models in the RW metric by considering the various implications of TD relationships~\mbox{\cite{Lee:2020zts,Lee:2023FoP,Lee:2022heb,SLee:2024}.} Numerous endeavors have aimed to measure TD. One approach involves directly observing TD through the decay time of distant supernova (SN) light curves and spectra~\cite{Leibundgut:1996qm,SupernovaSearchTeam:1997gem,Foley:2005qu,Blondin:2007ua,Blondin:2008mz}. Another method entails measuring TD by investigating the stretching of peak-to-peak timescales in gamma-ray burst (GRB) data~\cite{Norris:1993hda,Wijers:1994qf,Band:1994ee,Meszaros:1995gj,Lee:1996zu,Chang:2001fy,Crawford:2009be,Zhang:2013yna,Singh:2021jgr}. Efforts have also been made to detect the TD effect in the light curves of cosmologically distant quasars (QSOs)~\cite{Hawkins:2001be,Dai:2012wp}.  However, to date, no compelling evidence has emerged for cosmic time dilation, with conflicting results from various measurements. In the absence of explicit TD laws, the RW metric's speed of light may vary with cosmological time, akin to other physical properties such as mass density, temperature, and fundamental constants such as the Planck constant~\cite{Lee:2022heb}. This variability suggests a plausible scenario known as VSL with cosmic time. When characterizing the Friedmann--Lema\^itre--Robertson--Walker (FLRW) universe background, a hypersurface of constant time can be delineated based on physical quantities such as temperature or density, owing to the Universe's homogeneity, which ensures uniform temperature and density at each cosmic time. Nevertheless, it is vital to acknowledge that temperature and mass density may experience redshift due to the Universe's~expansion.

Relying solely on the cosmological principle, the method of cosmography serves as a kinematic description of the Universe's evolution, specifically focusing on the dynamics of cosmological expansion. Cosmography provides a versatile framework for handling cosmological parameters, relying solely on the kinematics of the Universe. This model-independent approach eliminates the need for dependence on any specific theoretical model, enabling a more generalized analysis. Notably, it abstains from defining any model a priori, thereby avoiding the necessity to postulate gravity for determining the Universe’s dynamics. The effectiveness of cosmography is demonstrated by its ability to choose models consistent with cosmological observations. Cosmography primarily concentrates on the later phases of the Universe’s development. The Taylor expansions associated with cosmography typically pertain to the observable domain where $z \ll 1$, facilitating the establishment of constraints on the current Universe. In our discussion, we explore the modification of late-time cosmography to apply the varying speed-of-light models. 

As is evident from the above equations, the effects of the meVSL model can manifest in each cosmographic parameter. It can potentially influence the values of the density contrasts in the $\Lambda$CDM model. In the subsequent follow-up paper, we will discuss the impact of this varying speed-of-light theoretical model on the values of density contrasts, comparing it with the standard model using several cosmological observational datasets. In the following chapter, we briefly introduce the meVSL model~\cite{Lee:2020zts,Lee:2023rqv,Lee:2023ucu,Lee:2023FoP,Lee:2022heb,SLee:2024}.  In Section \ref{Sec:Cos}, we compare the representations of cosmography in the meVSL model for each observable with those in the standard model, highlighting differences.  In Section \ref{Sec:Obs}  , we investigate the luminosity distance expression within the meVSL model's framework. Section \ref{Sec:Par}  describes how cosmographic parameters are represented by density contrasts when applying the LambdaCDM model. The final section concludes with a discussion.

\section{Brief Review of the Minimally Extended Varying Speed-of-Light (meVSL) Model}
\label{Sec:meVSL}

In this session, we will briefly review the potential of the VSL model and introduce the meVSL model~\cite{Lee:2020zts,Lee:2023rqv,Lee:2023ucu,Lee:2023FoP,Lee:2022heb,SLee:2024}. One can describe the homogeneous and isotropic spacetime at a specific time $t_l$ as 
\begin{align}
ds_{l}^2 = - c_l^2 dt^2 + a_l^2 \left[ \frac{dr^2}{1-Kr^2} + r^2  \left( d \theta^2 + \sin^2 \theta d \phi^2 \right)  \right] \quad \textrm{at} \,\,  t= t_l \label{dststar} \,. 
\end{align}

By incorporating Weyl’s postulate to extend the metric described in Equation~\eqref{dststar} to cosmic time $t$, we can formulate the line element as 
\begin{align}
ds^2 = - c(t)^2 dt^2 + a(t)^2 \left[ \frac{dr^2}{1-Kr^2} + r^2  \left( d \theta^2 + \sin^2 \theta d \phi^2 \right)  \right] \equiv - c(t)^2 dt^2 + a(t)^2 dl_{3\textrm{D}}^2 \label{dstgen} \,. 
\end{align} 

\begin{figure}[H]
 \includegraphics[width=0.9\textwidth]{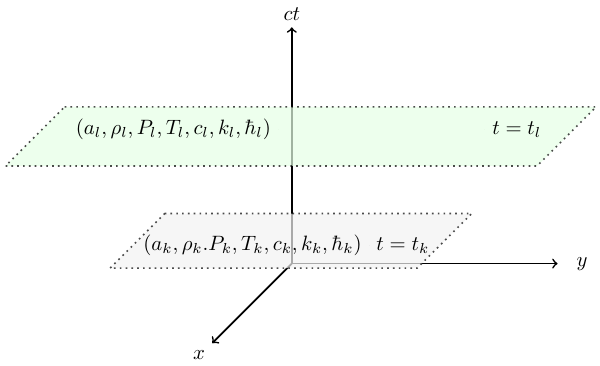} 
	\caption{At $t = t_k$, the values of physical quantities and constants, such as $a_k$, $\rho_k$, $P_k$,  $T_k$, $c_k$,  $k_{k}$, and $\hbar_k$, are fixed and independent of spatial position on the $t=t_k$ hypersurface. As the universe expands, these quantities and constants transition to $a_l$, $\rho_l$, $P_l$,  $T_l$, $c_l$,  $k_l$, and $\hbar_l$. The CP and Weyl’s postulate do not restrict $c_k$ to be equal to $c_l$; its value is determined by the cosmological TD relation. }
	\label{Fig1}
 
\end{figure}

 In this equation, the speed of light is expressed as a function of time, deviating from the conventional RW metric. Initially, this formulation may appear unconventional or counterintuitive. However, as depicted in Figure~\ref{Fig1}, the original RW metric implies that on hypersurfaces defined by $t_l$ or $t_k =$ constants, various quantities such as the scale factor $a_l = a(t_l)$, mass density $\rho_l = \rho(t_l)$,  pressure $P_l = P(t_l)$, temperature $T_l = T(t_l)$, speed of light $c_l = c(t_l)$,  Boltzmann constant $k_{l} = k(t_l)$, and Planck constant $\hbar_l = \hbar(t_l)$ remain constant regardless of the 3D spatial position. However, according to Weyl’s postulate, these quantities or constants can be expressed as functions of cosmic time $t$, accounting for cosmological redshift, as depicted in Figure~\ref{Fig1}. Traditionally, it has been assumed that physical constants, including the speed of light, remain constant over cosmic time. This additional assumption, namely that the speed of light remains constant ($c_l = c_k$) regardless of cosmic time, is not directly linked to the two conditions necessary to derive the RW metric: the CP and Weyl’s postulate. The constancy of the speed of light relies on cosmological time dilation, and it is essential to note that the GR does not specify any particular physical laws governing this constancy, as we will elucidate shortly. As the Universe progresses from $t_k$ to $t_l$, physical quantities such as $a(t)$, $\rho(t)$, $P(t)$, and $T(t)$ change over cosmic time $t$. The precise functional expressions for these quantities are obtained through the solution of Einstein’s Field Equations (EFEs) and Bianchi’s identity (BI), considering the equation of the state of fluids~\cite{Lee:2020zts,Lee:2023FoP}.

\vspace{-6pt}

\subsection{Cosmological Redshift}
\label{subsec:CRedshift}

One can rewrite the RW metric in Equation~\eqref{dstgen} as
\begin{align}
ds^2 = -(d X^0)^2 +  a^2(t) \left( \frac{dr^2}{1-kr^2} + r^2 \left( d \theta^2 + \sin^2 \theta d \phi^2 \right) \right) \equiv -(d X^0)^2 +  a^2(t) dl_{3\textrm{D}}^2 \label{RW} \,,
\end{align}
where $X^0 = ct$.  In this metric, the light signal propagates along the null geodesic $ds^2 = 0$, and, from metric \eqref{RW}, we obtain the outgoing light signals
\begin{align}
dl_{3\textrm{D}}(r\,, \theta\,, \phi) = \frac{dX^0}{a(t)} \label{dl3D} \,.
\end{align}

The spatial infinitesimal line element $dl_{3\textrm{D}}$ is a function of the comoving coordinates ($\sigma\,,\theta\,,\phi$) only and thus should be the same value at any given time.  From this fact, the standard model of cosmology (SMC, {i.e.}, by assuming a constant speed of light) obtains the cosmological redshift relation 
\begin{align}
\frac{dX^0(t_1)}{a(t_1)} = \frac{dX^0(t_2)}{a(t_2)}\quad \Rightarrow \quad \frac{dt_1}{a_1} = \frac{dt_2}{a_2} \quad \Rightarrow \quad \lambda_1 = c dt_1 \equiv c \nu_1^{-1} = \frac{a_1}{a_2} \lambda_2 \label{z} \,,
\end{align}
where $d t_i$ is the time interval of successive crests of light at $t_i$ ({i.e.},  the inverse of the frequency $\nu_i$ at $t_i$)~\cite{Weinberg:2008}. 
Traditionally, we obtain the cosmological redshift under the assumption that the speed of light is constant at these scales.  However, the Lorentz invariance (LI) is a local symmetry that is only meaningful at each spacetime point (event), while the GR is valid at cosmological scales. Therefore, the quibble about whether SR is generally adaptable at cosmological distances and time scales should be determined by observations~\cite{Roos15}. In this case, we can rewrite Equation~\eqref{z} as
\begin{align}
\frac{dX^0(t_1)}{a(t_1)}  = \frac{dX^0(t_2)}{a(t_2)} \quad \Rightarrow \quad \frac{c_1 dt_1}{a_1} = \frac{c_2 dt_2}{a_2} \quad \Rightarrow \quad \lambda_1 = c_1 dt_1 = \frac{a_1}{a_2} \lambda_2 \label{zVSL} \,,
\end{align}
where $c_i \equiv c (t_i)$, $a_i \equiv a(t_i)$, and 
\begin{align}
dX^{0} = d \left(\tilde{c} t \right) = \left(\frac{d \ln \tilde{c}}{d \ln t} + 1 \right) \tilde{c} dt \equiv c dt \quad \textrm{and} \quad \delta c \equiv \frac{c}{\tilde{c}} = \left(\frac{d \ln \tilde{c}}{d \ln t} + 1 \right) \label{dX0} \,,
\end{align}

Thus, the cosmological redshift relation still holds even if we allow the speed of light to vary as a function of cosmic time.  These results have been derived before in a so-called meVSL model to satisfy Einstein's field equations~\cite{Lee:2020zts}.

\subsection{The Possibility of Varying Speed-of-Light Theory in the Robertson--Walker Metric}
\label{subsec:FLRW}

The derivation of redshift involves employing the geodesic equation for a light wave, where $ds^2 = 0$ as in 
Equation~\eqref{dstgen}. The consistency of $dl_{3\textrm{D}}$ over time is ensured by the exclusive use of comoving coordinates. Expanding upon this groundwork, we reach the expression for outgoing light signals as
\begin{align}
d l_{3\textrm{D}} &= \frac{c(t_i) dt_i}{a(t_i)} \quad : \quad \frac{c_1 dt_1}{a_1} = \frac{c_2 dt_2}{a_2} \Rightarrow \begin{cases} c_1 = c_2 = c & \textrm{if} \quad \frac{dt_1}{a_1} = \frac{dt_2}{a_2} \qquad \textrm{SMC} \\ 
 c_1 = \frac{f(a_2)}{f(a_1)} \frac{a_1}{a_2} c_2 & \textrm{if} \quad \frac{dt_1}{f(a_1)} = \frac{dt_2}{f(a_2)} \quad \textrm{VSL} \\ c_1 = \left( \frac{a_1}{a_2}\right)^{\frac{b}{4}} c_2 & \textrm{if} \quad \frac{dt_1}{a_1^{1-\frac{b}{4}}} = \frac{dt_2}{a_2^{1-\frac{b}{4}}} \quad \textrm{meVSL}  \end{cases} \,, \label{dl3D}
\end{align}
where  $d t_i = 1/ \nu(t_i)$ represents the time interval between successive crests of light at $t_i$ ({i.e.}, the inverse of the frequency $\nu_i$ at $t_i$) and $f(a_i)$ denotes an arbitrary function of $a(t_i)$~\cite{Weinberg:2008}.

 {In the SMC, an additional assumption is introduced, positing that the cosmological TD between two hypersurfaces at $t = t_1$ and $t = t_2$ is proportional to the inverse of the scale factors $a(t)$ at each respective time. This assumption leads to the conclusion that the speed of light on these hypersurfaces remains constant, $c(t_1) = c(t_2) = c$. However, this assumption lacks derivation from any physical laws. Moreover, within the framework of GR, there exists no inherent physical basis for the constancy of the speed of light across cosmic time, as it holds solely for the local inertial observer. }

Conversely, in an expanding universe, the progression from one hypersurface to another results in an increase in the scale factor, naturally leading to the cosmological redshift of various physical quantities, including mass density and temperature. However, it is impossible to conclude about cosmological TD based solely on the CP and Weyl’s postulate in the RW metric. Instead, establishing such relationships relies on experimental observations. Efforts to measure cosmological time dilation have included direct observations of SN light curves and spectra to evaluate decay times of distance~\cite{Lee:2023ucu,Leibundgut:1996qm,SupernovaSearchTeam:1997gem,Foley:2005qu,Blondin:2008mz}. Another avenue to explore cosmological TD involves analyzing the elongation of peak-to-peak timescales observed in GRBs~\cite{Norris:1993hda,Wijers:1994qf,Band:1994ee,Meszaros:1995gj,Lee:1996zu,Chang:2001fy,Crawford:2009be,Zhang:2013yna,Singh:2021jgr}. Additionally, researchers have investigated TD effects within the light curves of cosmologically distant QSOs~\cite{Hawkins:2001be,Dai:2012wp,Lewis:2023jab}.  However, current observational evidence does not definitively confirm an exact correspondence between cosmological TD and predictions made by the SMC. Moreover, the RW model lacks a mechanism to determine cosmological TD conclusively. Thus, it remains valuable to explore the possibility of VSL in these observations, provided that the findings are consistent with those predicted by the SMC.

\subsection{The Modification of Einstein's Field Equations}
\label{subsec:EFE}

{In the meVSL, the line element is written as
\begin{align}
ds^2 = - c^2 dt^2 + a^2 \gamma_{\ij} dx^{i} dx^{j} \label{dsFRW} \,.
\end{align}

Then, the Riemann curvature tensors, Ricci tensors, and Ricci scalar curvature are given by \vspace{-15pt}

\begin{align}
&\tensor{R}{^0_i_0_j} = \frac{g_{ij}}{c^2} \left( \frac{\ddot{a}}{a} - H^2 \frac{d \ln c}{d \ln a} \right) , \quad
\tensor{R}{^i_0_0_j} = \frac{\delta^{i}_{j}}{c^2} \left( \frac{\ddot{a}}{a} - H^2 \frac{d \ln c}{d \ln a} \right) , \nonumber \\
&\tensor{R}{^i_j_k_m} =  \left( \frac{H^2}{c^2} + \frac{k}{a^2} \right) \left( \delta^{i}_{k} g_{jm} - \delta^{i}_{m} g_{jk} \right) \label{tRijkmApp} \,, \\
& R_{00} = - \frac{3}{c^2} \left( \frac{\ddot{a}}{a} - H^2 \frac{d \ln c}{d \ln a} \right)  \quad , \quad
R_{ii} = \frac{g_{ii}}{c^2} \left( 2 \frac{\dot{a}^2}{a^2} + \frac{\ddot{a}}{a} + 2 k \frac{c^2}{a^2} - H^2 \frac{d \ln c}{d \ln a} \right)
 \label{tR00mp} \,, \\
& R = \frac{6}{c^2} \left( \frac{\ddot{a}}{a} + \frac{\dot{a}^2}{a^2} + k \frac{c^2}{a^2} - H^2 \frac{d \ln c}{d \ln a} \right) \label{tRmp} \,.
\end{align} 

The energy-momentum tensor is given by
\begin{align}
T_{\mu}^{\nu} = \text{diag} \left(-\rho c^2, P, P, P \right) \label{Tmunump} \,.			
\end{align} 

One needs to investigate Bianchi's identity to provide the energy conservation given by 
\begin{align}
&\rho_i c^{2} = \rho_{i0} c_0^2 a^{-3 (1 + \omega_i)} \label{rhotc2mp} \,, 
\end{align}
where $c_0$ is the present value of the speed of light, $\rho_{i0}$ is the present value of mass density of the i-component, and we use $a_0 = 1$.  We obtain EFEs including the cosmological constant by using Equations~\eqref{tR00mp}--\eqref{rhotc2mp}
\begin{align}
& \frac{\dot{a}^2}{a^2} + k \frac{c^2}{a^2}  -\frac{ \Lambda c^2}{3} = \frac{8 \pi G}{3} \sum_{i} \rho_i \label{tG00mp} \,, \\ 
& \frac{\dot{a}^2}{a^2} + 2 \frac{\ddot{a}}{a} +  k \frac{c^2}{a^2} - \Lambda c^2 - 2 H^2 \frac{d \ln c}{d \ln a}  = -8 \pi G \sum_{i} \frac{P_i}{c^2} = -8 \pi G \sum_{i} \omega_{i} \rho_{i} \label{tG11mp} \,. 
\end{align}

One subtracts Equation~\eqref{tG00mp} from Equation~\eqref{tG11mp} to obtain 
\begin{align}
& \frac{\ddot{a}}{a} = -\frac{4\pi G}{3} \sum_{i} \left( 1 + 3 \omega_i \right) \rho_i  + \frac{\Lambda c^2}{3} + H^2 \frac{d \ln c}{d \ln a} \label{tG11mG00mp} \,. 
\end{align} 

From Equations~\eqref{tG00mp} and \eqref{tG11mG00mp}, one can understand that the expansion velocity of the Universe depends on not only the speed of light $c$ but also both on $G$  and on $\rho$. Additionally, so does the acceleration of the expansion of the Universe. 

Equation~\eqref{tG11mG00mp} also should be obtained by differentiating Equation~\eqref{tG00mp} with respect to the cosmic time $t$ and using Equation~\eqref{rhotc2mp}. This provides the relation between $G$ and $c$ as 
\begin{align}
\frac{d \ln G}{d \ln a} = 4 \frac{d \ln c}{d \ln a} \equiv b = \text{const.} \quad \Longrightarrow \quad \frac{G}{G_0} = \left( \frac{c}{c_0} \right)^4 = \left( \frac{a}{a_0} \right)^{b} = a^{b} \label{dotGoGmp} \,.
\end{align}

From the above Equations~\eqref{dotGoGmp}, one can obtain the expressions for the time variations of $c$ and $G$ as
\begin{align}
 \frac{\dot{G}}{G} = b H \quad , \quad \frac{\dot{c}}{c} = \frac{b}{4} H \label{dotcoc} \,. 
\end{align}

Thus, both the ratio of the time variation in the gravitational constant to the gravitational constant at the present epoch and the ratio of the time variation in the speed of light to the speed of light at the present epoch are given by 
\begin{align}
 \frac{\dot{G}_0}{G_0} = b H_0 \quad , \quad \frac{\dot{c}_0}{c_0} = \frac{b}{4} H_0 \label{dotcoc0} \,. 
\end{align} 

For more detailed information, please refer to Reference~\cite{Lee:2020zts}.

In the meVSL model, we consider local thermodynamics, energy conservation, and other local physics. These considerations induce the time evolutions of other physical constants and quantities shown in Table~\ref{tab:table-1}.

\begin{table}[htbp]
\caption{Summary for cosmological evolutions of physical constants and quantities of the meVSL model. These relations satisfy all known local physics laws, including special relativity, thermodynamics, and electromagnetic force \cite{Lee:2020zts}.}
\label{tab:table-1}
\begin{adjustbox}{width=\columnwidth,center}
\begin{tabular}{|c||c|c|c|}
	\hline
	local physics laws & Special Relativity & Electromagnetism & Thermodynamics \\
	\hline \hline
	quantities & $m = m_0 a^{-b/2}$ & $e = e_0 a^{-b/4}\,, \lambda = \lambda_0 a \,, \nu = \nu_0 a^{-1+b/4}$ & $T = T_0 a^{-1}$ \\
	\hline
	constants & $c = c_0 a^{b/4} \,, G = G_0 a^{b}$ & $\epsilon = \epsilon_0 a^{-b/4} \,, mu = \mu_0 a^{-b/4} $ & $k_{\textrm{B} 0} \,, \hbar = \hbar_0 a^{-b/4}$ \\
	\hline
	energies & $m c^2 = m_0 c_0^2$ & $h \nu = h_0 \nu_0 a^{-1}$ & $k_{\textrm{B}} T = k_{\textrm{B}} T_0 a^{-1}$ \\
	\hline
\end{tabular}
\end{adjustbox}
\end{table}

\section{Cosmolography of Varying Speed-of-Light Models}
\label{Sec:Cos} 

As seen in the previous session, due to the limited observational methods available to delineate the VSL models, it is crucial to identify observational methods that can be compared with the SMC. One such method we explore in this section 
is cosmography.
The Friedmann--Lema\^{i}tre--Robertson--Walker (FLRW) metric of varying speed-of-light (VSL) models is given by
\begin{align}
ds^2 = -c(t)^2 dt^2 + a(t)^2 \left[ \frac{dr^2}{1 - kr^2} + r^2 d \theta^2 + r^2 \sin^2 \theta d \phi^2 \right] \label{ds2} \,,
\end{align}
where $k$ is a constant related to curvature of space and $a(t)$ is the scale factor describing the evolution of the Universe. 
The coordinates $(t,r,\theta,\phi)$ of the RW metric are called comoving coordinates.  The spatial coordinates of objects remain the same, but their physical (proper) distance grows with time as space expands. The cosmic time gives the time a comoving observer measures at $(r,\theta,\phi) =$ constant.  Let the photon be emitted at a time $t_{\ast}$ with $r = 0$ and absorbed by an observer located at $r = r_0$ at the present epoch $t_0$.  One can express this time as $t_{\ast} = t_0 - \delta t$, where the time difference $\delta t$ is significantly smaller than the current cosmic time $t_0$. This relationship implies that the emitting galaxy is close to our location, and the time it takes for the photon to reach us is negligible compared to the current cosmic time. Thus, one can Taylor expand the scale factor about $t_0$ \vspace{-12pt}
\begin{align}
a(t_{\ast}) &= a(t_0) + \sum_{n=1}^{\infty} \frac{1}{n!} \frac{d^n a(t)}{d t^n} \Bigg |_{t=t_0} \left(t_{\ast} - t_0 \right)= a_0 - \dot{a}_0 \delta t + \frac{1}{2!} \ddot{a}_0  \delta t^2 - \frac{1}{3!} \dddot{a}_0  \delta t^3 + \frac{1}{4!} a^{(4)}_0  \delta t^4 + \cdots \nonumber \\ 
&\equiv 1 - H_0 \delta t - \frac{1}{2} q_0 H_0^2 \delta t^2 - \frac{1}{3!} j_0 H_0^3 \delta t^3 + \frac{1}{4!} s_0 H_0^4  \delta t^4 + \cdots \label{at} \,,
\end{align}
where $a(t_0) \equiv a_0 = 1$ denotes the value of the scale factor at the present epoch.  The so-called cosmographic parameters $H_0$, $q_0$, $j_0$, and $s_0$ are the Hubble, deceleration, jerk, and snap parameters, respectively.  They are dimensionless. In cosmology, the Hubble, deceleration, jerk, and snap parameters characterize the first four-time derivatives of the scale factor. These parameters elucidate the rate and acceleration/deceleration of cosmic expansion, serving as valuable tools for comprehending the behavior and expansion history of the Universe. Widely employed in observational data analysis and cosmological research, they aid in probing cosmological models and measuring significant quantities such as the density of matter in the Universe and the equation of state for dark energy. While the scale factor $a$ is an unobservable quantity, the cosmological redshift $z$ is an observable physical quantity. Moreover, one can also define the cosmological redshift in the vicinity of small values for $\delta z = z_{\ast}-z_0 = z_{\ast}$ using these parameters as
\begin{align}
z_{\ast} &= \frac{a_0}{a(t_0 - \delta t)} -1 \simeq  H_0 \delta t + \frac{1}{2!} \left( 2 +q_0 \right) \left( H_0 \delta t\right)^2 + \frac{1}{3!} \left( 6 + 6 q_0 + j_0 \right) \left( H_0 \delta t\right)^3 \nonumber \\&+ \frac{1}{4!}  \left( 24 + 36 q_0 + 6 q_0^2 + 8 j_0 - s_0 \right) \left( H_0 \delta t\right)^4 + \cdots  \label{zdt} \,,
\end{align}
where $z_0$ is the present cosmological redshift and equals $0$. 

Now, let us apply cosmographic parameters to observable quantities, starting with the Hubble parameter.  The Hubble parameter in the Taylor series is  \vspace{-12pt}
\begin{align}
H(\delta z) &\simeq H(z_0) + \frac{dH}{dz} \Bigg |_{z_0} \delta z + \frac{1}{2!} \frac{d^2H}{dz^2} \Bigg |_{z_0} \delta z^2 + \frac{1}{3!} \frac{d^3H}{dz^3} \Bigg |_{z_0} \delta z^3 + \cdots \nonumber \\
&\simeq H_0 \left[ 1 + \left( 1 + q_0 \right) \delta z + \frac{1}{2!} \left( j_0 - q_0^2 \right) \delta z^2 + \frac{1}{3!} \left( 3 q_0^2 + 3 q_0^3 - 4 q_0 j_0 - 3j_0 -s_0 \right) \delta z^3 + \cdots \right] \label{Hdeltaz} \,.
\end{align} 

This result holds consistently for both the standard cosmology model (SCM) and the VSL models.

Considering a null geodesic in an FLRW metric along the line of sight (LOS), the total LOS comoving distance can be obtained using Equation~\eqref{ds2}
\begin{align}
D_c(r_0) &\equiv \int_{0}^{r_0} \frac{dr}{\sqrt{1 - kr^2}} \equiv S_{k}^{-1}[r_0] = \left\{\begin{array}{cr} \frac{1}{\sqrt{k}} \sin^{-1} \left[ \sqrt{k} r_0 \right] &, \, k > 0 \\ r_0 &, \, k = 0 \\ \frac{1}{\sqrt{|k|}} \sinh^{-1} \left[ \sqrt{|k|} r_0 \right] &,\, k < 0 \end{array} \right\} \nonumber \\ &= \int_{t_{\ast}}^{t_0} \frac{c(t)}{a(t)} dt = \int_{z_0}^{z_{\ast}} \frac{c(z)}{H(z)} dz \label{fr0} \,.
\end{align}

Notably, in contrast to the SCM, the comoving distance $D_c$ is influenced by the $z$-dependent speed of light. Consequently, the transverse comoving distance $r_0$ can be derived from Equation \eqref{fr0}
\begin{align}
r_0(z_{\ast}) \equiv S_{k}[D_c(r_0)] =  \left\{\begin{array}{cr} \frac{1}{\sqrt{k}} \sin \left[ \sqrt{k}  \int_{z_0}^{z_{\ast}} \frac{c dz}{H(z)} \right] &, \, k > 0 \\  \int_{z_0}^{z_{\ast}} \frac{c dz}{H(z)} = D_c &, \, k = 0 \\ \frac{1}{\sqrt{|k|}} \sinh \left[ \sqrt{|k|}  \int_{z_0}^{z_{\ast}} \frac{c dz}{H(z)} \right] &,\, k < 0  \end{array} \right.  \label{r0z} \,.
\end{align}

Given that the acceleration of the Universe is a relatively recent phenomenon, we can focus our analysis on the vicinity of small values for the redshift interval $\delta z = z_{\ast}-z_0 = z_{\ast}$ in Equation \eqref{r0z}. For a short redshift interval $\delta z \equiv z_{\ast} \simeq 0$, one can expand the Taylor series of $r_0$ around $z_0$\vspace{6pt}

 \begin{align}
r_0(\delta z) \simeq  \left\{\begin{array}{cr} \int^{z_{0} + \delta z}_{z_0} \frac{c(z) dz}{H(z)} - \frac{k}{3 !}  \left( \int^{z_{0}+ \delta z}_{z_0} \frac{c(z) dz}{H(z)} \right)^3 &, \, k > 0 \\ \int^{z_0+\delta z}_{z_0} \frac{c(z) dz}{H(z)} = r_0 &, \, k = 0 \\ \int^{z_{0}+\delta z}_{z_0} \frac{c(z) dz}{H(z)} - \frac{k}{3 !}  \left( \int^{z_{0}+\delta z}_{z_0} \frac{c(z) dz}{H(z)} \right)^3 &,\, k < 0 \end{array} \right. \label{r0zapp} \,.
\end{align}

Measuring distances to cosmological objects stands as the primary method for probing the cosmic metric and deciphering the expansion history of the Universe. In the meVSL model, the comoving distance at redshift $z_{\ast}$ is given by  
\begin{align}
 D_{c}(r_0) &= \int^{z_{0}+\delta z}_{z_0} \frac{c(z) dz}{H(z)} \simeq  \int^{z_{0}+\delta z}_{z_0} \frac{c_0}{H_0} dz \left[ 1 + B_0 \delta z + C_0 \delta z^2 + D_0 \delta z^3 + \cdots \right] \nonumber \\ &= \frac{c_0}{H_0} \delta z  \left[ 1 + \frac{B_0}{2} \delta z + \frac{C_0}{3} \delta z^2 + \frac{D_0}{4} \delta z^3 + \cdots \right] \label{intzodz} \,,
\end{align}
where\vspace{-12pt}

\begin{align}
B_0 &= - \frac{H'_0}{H_0} + \frac{c_0'}{c_0} = -\left(q_0 + 1 + \frac{b}{4} \right) \label{B0} \,, \\
C_0 &= \frac{1}{2!} \left[ -  \frac{H''_0}{H_0}  + 2 \left( \frac{H'_0}{H_0} \right)^2 + \frac{c''_0}{c_0} -2 \frac{c'_0}{c_0} \frac{H'_0}{H_0}\right] = \frac{1}{2} \left[ -j_0 + 3q_0^2 + 4q_0 +2 +\frac{b}{4} \left( \frac{b}{4} + 2q_0 + 3 \right) \right] \label{C0} \,, \\
D_0 &= \frac{1}{3!} \left\{ -  \frac{H'''_0}{H_0} + 6 \frac{H''_0}{H_0} \frac{H'_0}{H_0}  - 6 \left( \frac{H'_0}{H_0} \right)^3 + \frac{c'''_0}{c_0} -3 \frac{c''_0}{c_0} \frac{H'_0}{H_0} -3 \frac{c'_0}{c_0} \left( \frac{H''_0}{H_0} - 2 \left( \frac{H'_0}{H_0} \right)^2  \right) \right\} \label{D0} \\
	&=  \frac{1}{3!} \left[ s_0 + 9 j_0 + 10 q_0 j_0 -15 q_0^3 - 27 q_0^2 -18 q_0 -6 + \frac{b}{4} \left( \left(3 j_0-9 q_0^2-15 q_0-11\right) - \frac{3}{4} (q_0+2) b - \frac{1}{16} b^2\right) \right]  \nonumber \,.
\end{align}

The transverse comoving distance, denoted as $r_0(\delta z)$, can be derived from\linebreak  {\mbox{Equations~\eqref{r0zapp} and~\eqref{intzodz}}}
\begin{align}
r_0(\delta z) \simeq \frac{c_0 \delta z}{H_0} \left[ 1 + \frac{1}{2!} B_0 \delta z + \frac{1}{3!} \left( 2 C_0 + \Omega_{k} \right) \delta z^2 + \frac{1}{4!}\left( 6 D_0 + 6 B_0 \Omega_{k} \right) \delta z^3 + \cdots \right] \label{r02z} \,.
\end{align}

The luminosity distance ($d_{L}$) for an object is expressed as $d_{L} = (L/4\pi F)^{1/2}$, where $L$ denotes the luminosity of the observed object (assumed to be known for high-redshift supernovae) and $F$ represents the energy flux received from the object. This formula describes the luminosity distance in a dynamic, homogeneous, isotropic spacetime. Expressing the luminosity distance as a function of $\delta z$ involves Equations~\eqref{zdt} and \eqref{r02z} \vspace{-12pt}

\begin{align}
d_{L}(\delta z) &= \left( 1 + \delta z \right) r_0 (\delta z)  \nonumber \\ &\simeq \frac{c_0 \delta z}{H_0} \left[ 1 + \frac{1}{2!} \left( 2 + B_0 \right) \delta z + \frac{1}{3!} \left(  3B_0 + 2C_0 + \Omega_{k} \right) \delta z^2 + \frac{1}{4!} \left( 8 C_0 + 6 D_0 + \left(4 + 6 B_0 \right) \Omega_k \right) \delta z^3 + \cdots \right] \nonumber \\ 
&\equiv \frac{D_{L}(\delta z)}{H_0} \label{DL} \,,
\end{align} 
which introduce 
the concept of the Hubble-free luminosity distance $D_L$, which remains independent of $H_0$. These results remain consistent with those found in references~\cite{Harrison:1976,Visser:2003vq,Dunsby:2015ers}, barring the impact of the time-varying speed of light. Consequently, by eliminating the derivation terms associated with $c$, the obtained results match those of the reference. Notably, our approach does not necessitate relying on lookback time to arrive at these~outcomes.

\section{Observation}
\label{Sec:Obs} 
The distance modulus ($\mu$) is the logarithmic measure of the ratio between an astronomical object's intrinsic and observed brightness, used for determining its distance. When dealing with standard candles such as Cepheid variables or Type Ia supernovae (SNe Ia), the distance modulus is widely used in astronomy to quantify the distance to celestial objects. It represents the difference between the apparent magnitude ({i.e.}, observed brightness),  $m(z)$ (preferably corrected for interstellar absorption effects), and the absolute magnitude ({i.e.}, intrinsic brightness), $M$, of a standard candle. The luminous distance measured in megaparsecs (Mpc), $d_{L}(z)$, is connected to the distance modulus through 
\begin{align}
\mu(z) = m(z) - M = 5 \log_{10} \left[ \frac{d_{L}(z)}{(\textrm{Mpc})} \right] + 25 \label{muz} \,.
\end{align} 

By replacing the luminosity distance in the above equation with the Hubble-free luminosity distance, $D_L(z)$, defined in Equation~\eqref{DL}, we obtain 
\begin{align}
\mu(z) = 5 \log_{10} \left[ \frac{D_{L}(z)}{ H_0 (\textrm{Mpc})} \right] + 25 = 5 \log_{10} \left[ \frac{D_{L}(z)}{(\textrm{km/s})}  \right] - 5 \log_{10} \left[ \frac{ H_0}{(\textrm{km/s/Mpc})} \right] + 25 \label{muz2} \,.
\end{align} 

Equations \eqref{muz} and \eqref{muz2} enable us to express the Hubble parameter in units of \linebreak (km/s/Mpc) as\vspace{-12pt}

\begin{align}
\log_{10} \left[ \frac{ H_0}{(\textrm{km/s/Mpc})} \right] = 0.2 M + \log_{10} \left[ \frac{D_{L}(z)}{(\textrm{km/s})}  \right]  -0.2 m(z) + 5 \equiv 0.2 M + a_B(z) + 5 \label{logH0} \,,
\end{align}
where the Hubble intercept parameter, denoted as $a_B$ and employed in the SH$0$ES (Supernova H$0$ for the Equation of State) analysis, defines the $x$-intercept ($m = 0$) of a Hubble diagram plotting $0.2 m$ against a modified $\log_{10} D_{L}$ term~\cite{Riess:2016jrr}. At low redshifts ($z \ll 1$), one can determine $a_B$ using Equations~\eqref{DL} and \eqref{logH0}}, 

\begin{align}
a_{B} \approx \log_{10} \left[  \frac{c z}{(\textrm{km/s})}  \left( 1 + \frac{1}{2!} \left( 2 + B_0 \right) z + \frac{1}{3!} \left(  3B_0 + 2C_0 + \Omega_{k} \right) z^2 + \frac{1}{4!} \left( 8 C_0 + 6 D_0 + \left(4 + 6 B_0 \right) \Omega_k \right) z^3 + \cdots \right) \right] -0.2 m(z) \label{aBVSL} \,,
\end{align}
where $a_B$ is measured from a group of SNe Ia with known redshifts and magnitudes $\left(z,m(0) \right)$.

\section{Parameters}
\label{Sec:Par} 
Cosmography proves to be a valuable tool in constraining cosmological parameters~\cite{Capozziello:2011tj,Muthukrishna:2016evq,Birrer:2018vtm,Rezaei:2020lfy}. The Hubble parameter in the meVSL model is derived from the modified Friedmann equation, incorporating the $\Lambda$CDM model~\cite{Lee:2020zts}
\begin{align}
& H^2 \equiv \left( \frac{\dot{a}}{a} \right)^2 = \left[ \frac{8 \pi G_0}{3} \rho_{\textrm{m}} + \frac{ \Lambda c_0^2}{3} - k \frac{c_0^2}{a^2} \right] a^{b/2} \label{HmeVSL} \,, \\
&1 = \left[ \frac{8 \pi G_0}{3H^2} \rho_{\textrm{m}} + \frac{ \Lambda c_0^2}{3H^2} - \frac{k c_0^2}{a^2H^2} \right] a^{b/2} \equiv \left[ \Omega_{\textrm{m}} +\Omega_{\Lambda} + \Omega_{k}  \right] a^{b/2} \label{OmegameVSL} \,,
\end{align}
where one defines the density contrasts of each component. Using these density contrasts, the coefficients in Equations~\eqref{B0}--\eqref{D0} can be expressed as 
\begin{align}
B_0 &= - \left( \frac{3}{2} \Omega_{\tm 0} + \Omega_{k0} \right) \label{B0LCDM} \,, \\
C_0 &= \frac{27}{4} \Omega_{\tm 0}^2 + 3 \left( 3 \Omega_{k0} - 1 \right) \Omega_{\tm0} + \left( 3 \Omega_{k0} - 1 \right) \Omega_{k0} \label{C0LCDM} \,, \\
D_0 &= -\frac{405}{8} \Omega_{\tm 0}^3 - \frac{81}{2} \left( \frac{5}{2} \Omega_{k0} - 1 \right) \Omega_{\tm0}^2 - \frac{3}{2} \left( 45 \Omega_{k0}^2 - 27  \Omega_{k0} + 2 \right)\Omega_{\tm0} - 3 \Omega_{k0}^2 \left(5 \Omega_{k0} -3 \right)  \label{D0LCDM} \,.
\end{align}

One can also use density contrasts of the $\Lambda$CDM model to express cosmographic parameters as

\begin{align}
q_0 &= \frac{3}{2} \Omega_{\tm 0} + \Omega_{k0} - 1 -\frac{b}{4} \label{q0LCDM} \,, \\
j_0 &= 1 - \Omega_{k0} + \frac{b}{8} \left[ b + 6 - 4 \left( 3 \Omega_{\tm 0} + 2 \Omega_{k 0} \right) \right] \label{j0LCDM} \,, \\
s_0 &= 1 + \left( \Omega_{k 0} + \frac{3}{2} \Omega_{\tm 0} - 2 \right) \Omega_{k 0} - \frac{9}{2} \Omega_{\tm 0} + \frac{b}{8} \left[ 12 + 8 \left( \Omega_{k 0} + 3 \Omega_{\tm 0} - 2 \right) \Omega_{k0} + 3\Omega_{\tm 0} \left( 6 \Omega_{\tm 0} -1 \right) \right] \nonumber \\ &+ \frac{b^2}{16} \left( 11 - 18 \Omega_{k 0} - 27 \Omega_{\tm 0} \right) + \frac{3}{32} b^3  \label{s0LCDM} \,.
\end{align}

As evident from the above equations, the effects of the meVSL model can manifest in each cosmographic parameter. It can potentially influence the values of the density contrasts in the $\Lambda$CDM model. In the subsequent follow-up paper, we will discuss the impact of this varying speed-of-light theoretical model on the values of density contrasts, comparing it with the standard model using several cosmological observational datasets.

\section{Discussion}

It has been well established that cosmography provides a model-independent means of constraining cosmological parameters. However, in this paper, we demonstrated that if we apply cosmography to a varying speed-of-light model, the predictions may differ from those offered by the standard cosmological model, depending on cosmological observables. Specifically, we have presented predictions for the minimally extended varying speed-of-light model in this paper. According to these predictions, cosmological parameters obtained from Hubble parameters or photometric distances can yield different values in the standard model and the meVSL model. We intend to delve deeper into this theoretical possibility through empirical investigations using specific cosmological observations.

\vspace{6pt}

\section*{Acknowledgments}
SL is supported by Basic Science Research Program through the National Research Foundation of Korea (NRF) funded by the Ministry of Science, ICT, and Future Planning (Grant No.  NRF-2017K1A4A3015188,  NRF-2019R1A6A1A10073079,  NRF-2022R1H1A2011281).  



\end{document}